\documentclass{article}
\usepackage{spconf,amsmath,graphicx,hyperref}
\usepackage{booktabs}
\usepackage{caption}
\usepackage{makecell}
\usepackage{array}

\title{EVALUATION OF PREPROCESSING PIPELINES IN THE CREATION OF IN-THE-WILD TTS DATASETS}
\name{%
  \begin{tabular}{c}
    Matías Di Bernardo$^{\star}$ \quad Emmanuel Misley$^{\star}$ \quad Ignacio Correa$^{\star}$ \\
    Mateo García Iacovelli$^{\star}$ \quad Simón Mellino$^{\star}$ \quad Gala Lucía Gonzalez Barrios$^{\dagger}$
  \end{tabular}%
}
  
  \address{$^{\star}$Universidad Nacional de Tres de Febrero \\
        matias.di.bernardo@hotmail.com \\
      $^{\dagger}$Virginia Tech \\
      gala@vt.edu}
      
\begin{document}
%
\maketitle
\begin{abstract}
This work introduces a reproducible, metric-driven methodology to evaluate preprocessing pipelines for \emph{in-the-wild} TTS corpora generation. We apply a custom low-cost pipeline to the first \emph{in-the-wild} Argentine Spanish collection and compare 24 pipeline configurations combining different denoising and quality filtering variants. Evaluation relies on complementary objective measures (PESQ, SI-SDR, SNR), acoustic descriptors (T$_{30}$, C$_{50}$), and speech-preservation metrics (F$_0$-STD, MCD). Results expose trade-offs between dataset size, signal quality, and voice preservation; where denoising variants with permissive filtering provide the best overall compromise for our testbed. The proposed methodology allows selecting pipeline configurations without training TTS models for each subset, accelerating and reducing the cost of preprocessing development for low-resource settings.
\end{abstract}
\section{Introduction}
\label{sec:intro}

Text-to-speech (TTS) technology has advanced rapidly and is now widely deployed across multimedia, communication, and assistive applications; modern modeling and training methods yield highly natural synthetic voices but remain strongly dependent on large volumes of high-quality recorded speech for training \cite{survey1}. Traditionally, such corpora are produced in controlled studio environments with careful phonetic design and strict quality assurance, which is costly and limits speaker and style diversity \cite{erica}. By contrast, \emph{in-the-wild} data (e.g., Internet-harvested or crowdsourced recordings) offer greater diversity, spontaneity and accent coverage and are therefore an attractive resource, especially for low-resource languages where professionally recorded material is scarce or prohibitively expensive \cite{tts_wild}.

The main challenge of \emph{in-the-wild} audio is high variability in recording conditions \cite{data2018}. Such recordings frequently contain background noise, reverberation, overlapping speech, and transcription errors, all of which degrade usability for TTS training unless mitigated by appropriate processing. To address these issues, the community has proposed a variety of automatic preprocessing pipelines that perform stages such as denoising, segmentation, speaker clustering, target-speaker extraction, and quality-based filtering and selection \cite{autoprep, emilia}. These frameworks have demonstrated that carefully designed data selection and cleaning can substantially enhance the utility of found audio data for TTS training \cite{improv_tts1}.

In recent years, several pipelines were introduced for TTS \cite{pipeline_tts1, pipeline_tts2}, for ASR \cite{pipeline_asr1, pipeline_asr2, pipeline_asr3} and for general dataset generation \cite{pipeline_data1, pipeline_data2}. Despite the growing number of pipeline configurations, the literature lacks systematic acoustic comparisons that quantify how individual preprocessing choices affect objective audio metrics. Different studies describe distinct curation strategies and provide application-level TTS results, but few report a comprehensive set of acoustic descriptors that would facilitate fair and reproducible comparisons between pipelines. This gap complicates the assessment of which pipeline components are most critical for obtaining studio-like data quality from wild recordings. Also, it serves as a baseline to contrast the effectiveness of different approaches in dataset curation \cite{alternaive1, alternative2}.

We contribute an open-source\footnote{https://github.com/MatiasDiBernardo/Lowcost-ITW-curation} and CPU-friendly preprocessing chain, with a reproducible methodology to assess preprocessing variants. Our design emphasizes simplicity and computational efficiency so that research groups with limited hardware can produce substantial, high-quality training material without requiring large GPU clusters. 


As a real-world case study, we apply the proposed pipeline to the creation of the first \emph{in-the-wild} Argentine Spanish corpus encompassing diverse regional accents. Existing Argentine Spanish resources are largely studio-recorded or limited in dialectal coverage \cite{google-arg, datset_arg}; to our knowledge, no public wild-harvested corpus exists that captures Argentina's accent variability. 

The main contributions of this paper are threefold: (i) we propose a reproducible methodology to evaluate and compare preprocessing pipelines independently of any specific TTS system, providing objective metrics to characterize pre/post processing effects; (ii) we develop a low-cost, CPU-friendly preprocessing chain designed to be practical and modular for research groups and communities working on low-resource languages; and (iii) we collect the first \emph{in-the-wild} Argentine Spanish corpus that captures regional dialectal diversity and serves as a real-world testbed for pipeline evaluation.

\section{Methodology}
\label{sec:methodology}

For \emph{in-the-wild} data, the primary goal of the preprocessing pipeline is to improve the quality conditions of the audio data. The main tools to achieve this are denoising or speech-enhancement algorithms and filtering based on non-intrusive quality assessment. Although prior pipelines report improvements in downstream TTS quality, it is difficult to identify a single “best” pipeline because datasets are rarely characterized both before and after processing.

To address this issue, we compute a set of complementary metrics that quantify different aspects of the corpus before (subscript $R$ for raw) and after processing (subscript $P$ for processed with a specific configuration). First, \textit{Dataset reduction} (DR) measures the relative loss of duration (Equation~\ref{eq:dr}); a smaller reduction is preferred. Second, \textit{Signal quality} (SQ) (Equation~\ref{eq:sq}) aggregates objective quality measures: PESQ and SI-SDR (computed with PyTorch Squim \cite{squim}) and SNR  (computed with WADA-SNR \cite{wada}). These metrics are expected to improve with respect to the raw dataset.

Next, \textit{Acoustic parameters} (AP) describe recording environment conditions, like energy distribution and reverberation (Equation~\ref{eq:ap}); T$_{30}$ is expected to decrease while C$_{50}$ is expected to increase. Both are computed with a CNN model validated for Argentine Spanish voices \cite{Maxi}. Finally, we establish \textit{Speech differences} (SD) as baseline prosodic and voice preservation metrics (Equation~\ref{eq:sd}), this includes any deviation of the original $F_0$ standard deviation (calculated with PESTO \cite{PESTO}) and the percentage increase in mean mel-cepstral distortion (MCD) \cite{mcd_metric} relative to an acceptable reference value of 5 dB \cite{survey2} (computed only for denoised audios).

\begin{subequations}\label{eq:subsets}
\begin{align}
DR_P &= 1 - \frac{\mathrm{HOURS}_P}{\mathrm{HOURS}_R}
\label{eq:dr} \\[4pt]
SQ_P &= 
\frac{\mathrm{PESQ}_R}{\mathrm{PESQ}_P}
\;+\; \frac{\mathrm{SI\text{-}SDR}_R}{\mathrm{SI\text{-}SDR}_P}
\;+\; \frac{\mathrm{SNR}_R}{\mathrm{SNR}_P}
\label{eq:sq} \\[4pt]
AP_P &= 
\frac{T_{30,P}}{T_{30,R}}
\;+\; \frac{C_{50,R}}{C_{50,P}}
\label{eq:ap} \\[4pt]
SD_P &= 
\left|1 - \frac{\mathrm{F0std}_P}{\mathrm{F0std}_R}\right|
\;+\; \frac{\mathrm{MCD}_P}{5}
\label{eq:sd}
\end{align}
\end{subequations}

We combine these subset scores into a single objective evaluated over pipeline configurations $P$ (Equation~\ref{eq:final}). In the default formulation, all subsets are equally weighted, though weight coefficients can be introduced to prioritize particular criteria.

\begin{equation}
\label{eq:final}
\min_{P \in \mathrm{Conf}}\;
\Big\{
DR_{P}
\;+\;
SQ_{P}
\;+\;
AP_{P}
\;+\;
SD_{P}
\Big\}
\end{equation}

These metrics are evaluated on the collected \emph{in-the-wild} Argentine Spanish dataset used in this study: 24 hours of audio from 59 speakers. The material was selected to maximize diversity in acoustic conditions and speech characteristics.

\section{PreProcessing pipeline}
\label{sec:pipeline}

\subsection{Voice activity detection (VAD)}


The first stage of our pipeline is voice activity detection (VAD), which removes non-speech segments and produces an initial segmentation of utterance boundaries. Following prior work \cite{emilia}, we adopt Silero VAD \cite{SileroVAD} as the baseline but introduce an adaptive hyperparameter optimization to handle speech rate variability, a common challenge in in-the-wild audio. Our method uses Whisper's timestamps to classify segments as slow, normal, or fast, and then applies a targeted Tree-Structured Parzen Estimator (TPE) optimization to find the ideal VAD settings for each category. Another key feature is the subsequent control over the final utterance length, where segments are concatenated to match a user-defined target mean and standard deviation. This capability is fundamental for adapting the preprocessed data to the specific input requirements of various downstream TTS models.



\subsection{Denoising and speech enhancement}

\begin{table*}[htb]
\centering
\caption{Dataset metrics for different pipeline stages for DeepFilterNet + NISQA: 3.8.}
\label{tab:dataset-quality-full}
\resizebox{\textwidth}{!}{%
\begin{tabular}{@{} l c
                  c c c
                  c c
                  c c @{}}
\toprule
\textbf{Dataset} & \textbf{Hours} &
\multicolumn{3}{c}{\textbf{Signal Quality}} &
\multicolumn{2}{c}{\textbf{Acoustic Parameters}} &
\multicolumn{2}{c}{\textbf{Speech differences}} \\

\cmidrule(lr){3-5}\cmidrule(lr){6-7}\cmidrule(l){8-9}
 & & \textbf{PESQ $\uparrow$} & \textbf{SNR $\uparrow$} & \textbf{SI-SDR $\uparrow$} & \textbf{T30 $\downarrow$} & \textbf{C50 $\uparrow$} & \textbf{F0 std} & \textbf{MCD} \\

\midrule
Original                     & $24.3$ & $2.82\pm0.72$  & $19.1\pm8.9$  & $17.8\pm6.9$  & $0.98\pm0.57$  & $15.9\pm5.5$  & $200.1\pm103.6$  & --- \\

Pipeline (no denoise)  & $5.1$ & $\mathbf{3.41\pm0.48}$ & $21.2\pm7.1$ &  $\mathbf{22.2\pm4.5}$ & $0.79\pm0.38$ & $17.9\pm4.1$  & $181.8\pm  82.82$ & ---        \\

Pipeline (denoised)    & $13.2$ & $3.28\pm0.49$ & $\mathbf{22.6\pm9.5}$ &  $21.1\pm4.8$ & $\mathbf{0.53\pm0.30}$ & $\mathbf{19.1\pm4.4}$ & $184.6 \pm  94.81$ & $2.79\pm2.34$ \\

Eliminated              & $19.2$ & $2.67\pm0.69$ &$18.5\pm9.3$ &  $16.7\pm7.0$ & $1.03\pm0.60$ & $15.3\pm5.7$ & $206.3 \pm  106.4$ & ---         \\
\bottomrule
\end{tabular}%
}
\end{table*}

The literature employs a wide range of models to improve the quality of \emph{in-the-wild} audio. Representative examples include Demucs \cite{tts_wild}, FRCRN/VoiceFixer \cite{pipeline_tts1}, and MBTFNet \cite{pipeline_tts2}; conversely, some pipelines used source separation as the only improvement of the audio, like Emilia \cite{emilia} that uses UVR-MDX-Net. These approaches differ in their objectives and operating points: some prioritize perceptual quality gains, while others emphasize signal fidelity or computational efficiency.

For the enhancement stage, it is particularly important to monitor the speech-difference metrics, since they quantify how denoising or restoration affects voice characteristics. Certain generative restoration methods can increase perceived quality yet also modify the original speaker timbre and prosody, producing audio that sounds more robotic, toneless, or broadcast-like. 


Because our design goal emphasizes low computational cost and portability, we avoid generative enhancement models that are computationally intensive or prone to altering speaker identity. Few state-of-the-art solutions run efficiently on CPU while remaining fast enough for large-scale processing. Under these constraints, we evaluate two practical denoising models that balance performance and efficiency: DeepFilterNet (DFN) \cite{deepfilter} and Demucs \cite{Demucs}.

\begin{figure}[tb]
  \centering
  \centerline{\includegraphics[width=8.5cm]{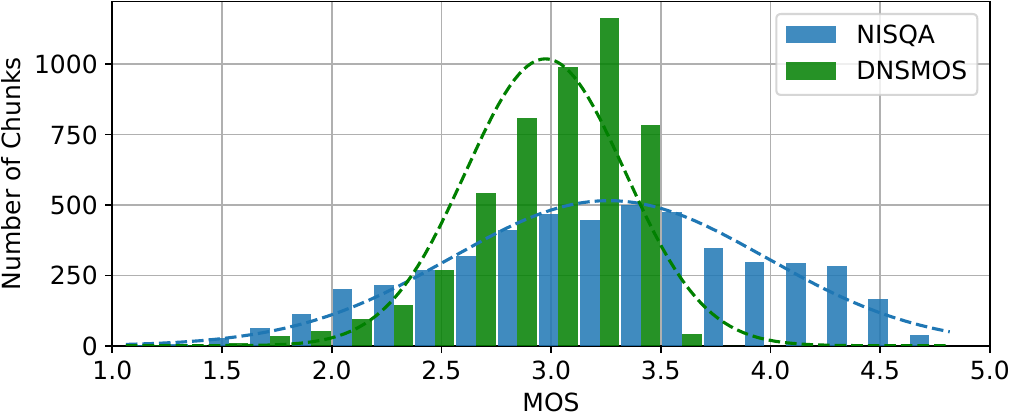}}
  \caption{MOS predicted value per chunk density.}
   \label{mos_testardo}
\end{figure}
\subsection{Quality filtering}

The vast majority of preprocessing pipelines employ non-intrusive quality assessment models to establish filtering thresholds, yet there is no unified criterion for which model or threshold to use.



To illustrate this issue, we compare NISQA \cite{nisqa} and DNSMOS \cite{dns_mos} quality scores computed over our raw dataset Figure \ref{mos_testardo}. This analysis highlights the challenge of comparing pipelines that rely on different non-intrusive metrics: DNSMOS shows a lower median and smaller variance compared to NISQA on our data. Consequently, small adjustments to a DNSMOS threshold can produce larger relative changes in the set of accepted utterances than equivalent adjustments to a NISQA threshold. This observation underscores the need for standardized evaluation practices or cross-metric analyses when reporting filtering decisions.

\subsection{Speech to text (STT)}
For transcription, the majority of the literature relies on Whisper Large \cite{whisper} due to its strong accuracy. Whisper Large, however, is computationally expensive and slow on CPU. In our experiments, alternative models that advertise faster CPU performance proved less reliable in terms of transcription quality. Because transcription correctness is central to producing a high-quality TTS corpus, we prioritize accuracy at this stage and accept the additional processing time.


\begin{figure}[tb]
  \centering
  \centerline{\includegraphics[width=8.5cm]{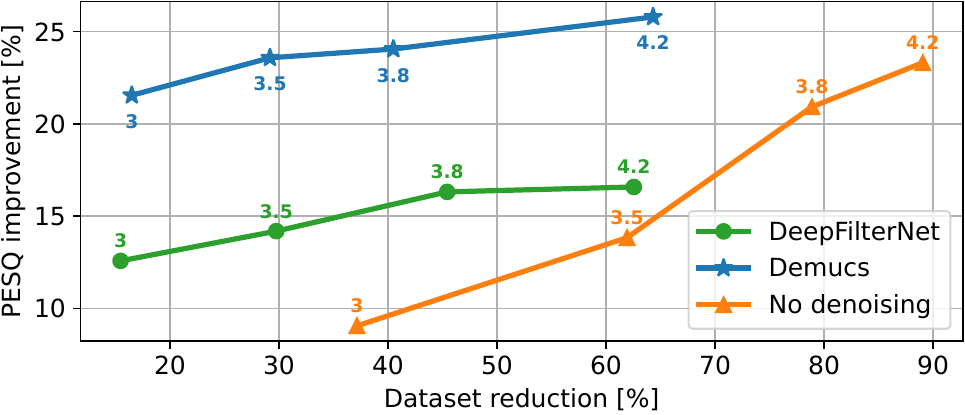}}
  \caption{Relation between dataset reduction and PESQ improvements with NISQA as filter.}
    \label{fig:pesq_vs_hs}
\end{figure}

\section{Experiments}
\label{sec:experimients}
We evaluated 24 different pipeline configurations. The corpus was processed under three denoising conditions (DeepFilterNet, Demucs, no-denoising) and filtered using NISQA and DNSMOS. Appropriate thresholds were chosen from the predicted-quality distributions to ensure equal utterance distribution: NISQA = \{3.0, 3.5, 3.8, 4.2\} and DNSMOS = \{2.7, 3.0, 3.2, 3.4\}.

Table~\ref{tab:dataset-quality-full} presents a stage-by-stage breakdown for one representative configuration (DeepFilterNet + NISQA, threshold = 3.8). After processing, all configurations show consistent metric improvements and lower standard deviations. This indicates higher and more uniform audio quality. The no-denoising variant achieves higher PESQ and SI-SDR for the retained subset, but keeps fewer hours. This illustrates a trade-off between quality and quantity. We computed metrics for all 24 variants to explore these trade-offs.

Figure~\ref{fig:pesq_vs_hs} compares dataset reduction and PESQ gains. Demucs yields the largest PESQ improvements across thresholds and the no-denoising variant shows the greatest PESQ gain under selective filtering. DeepFilterNet has lower PESQ improvement for a higher filter than no-denoising. Similar behavior appears in all signal-quality metrics where the Demucs variant consistently ranks higher. Denoised variants improve by no more than 8\% across filter conditions (NISQA, DNSMOS). No-denoising variants always improve by more than 10\%, but this comes with greater dataset reduction. A higher threshold always results in fewer but more uniform audio samples, consequently, it lowers the standard deviation for all metrics in every configuration.

\begin{figure}[tb]
  \centering
  \centerline{\includegraphics[width=8.5cm]{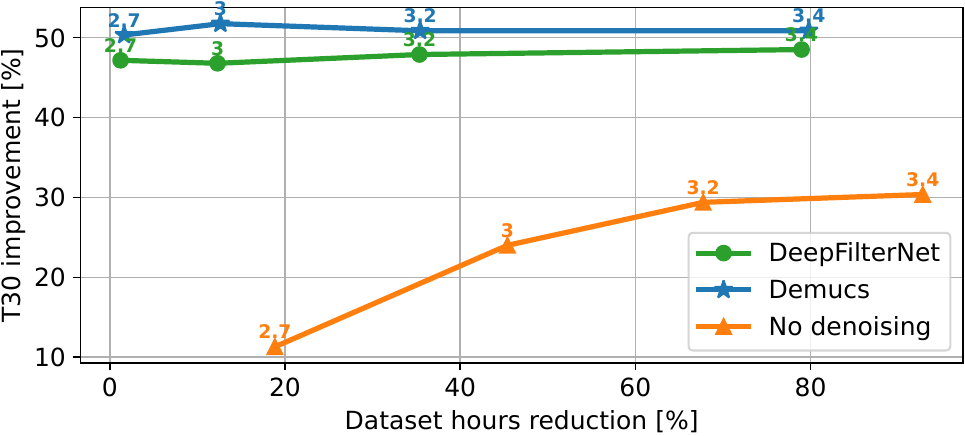}}
  \caption{Relation between dataset reduction and T30 improvements with DNSMOS as filter.}
    \label{fig:T30_dif}
\end{figure}

Acoustic parameters show similar trends. Filtering produces modest gains in denoised conditions, with less than 5\% difference. It gives considerable improvements for the no-denoising case (see Figure~\ref{fig:T30_dif}). There is no significant T$_{30}$ difference between DeepFilterNet and Demucs. However, Demucs has about a 5\% advantage on C$_{50}$.

\begin{figure}[tb]
  \centering
  \centerline{\includegraphics[width=8.5cm]{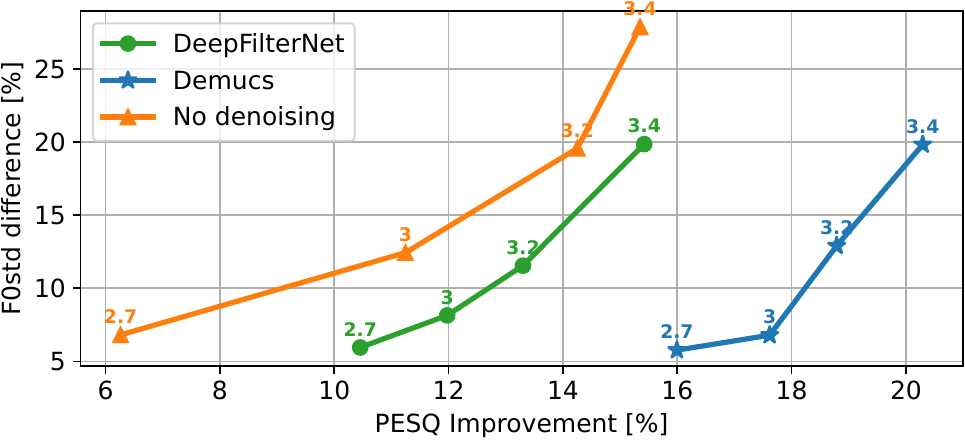}}
  \caption{Relation between F0std difference and PESQ improvements with DNSMOS as filter.}
    \label{fig:F0_vs_psq}
\end{figure}

Speech-difference metrics indicate a decrease in F$_0$ variability with stronger filtering, largely because poorer-quality speakers are removed. Denoisers that preserve voice timbre yield smaller F$_0$-STD changes as filtering becomes more aggressive (Figure~\ref{fig:F0_vs_psq}). MCD exhibits little change due to filtering; Demucs yields a slightly lower MCD than DeepFilterNet, while the no-denoising variant is not penalized in terms of MCD score.

Table~\ref{tab:config-scores} summarizes the proposed evaluation metrics for five representative configurations (ranked best to worst). Demucs variants achieve superior signal-quality and acoustic-parameter scores, while the no-denoising variants preserve speech characteristics best but perform worse on other criteria. We considered variance-normalization but chose not to apply it: the empirical variance of each metric across configurations reflects the extent to which preprocessing affects that measure, so metrics with larger variance provide more discriminative information for ranking configurations and therefore carry more weight in the composite score. 

For the evaluated dataset, the objective naturally favors configurations that minimize dataset reduction; given the relatively good initial conditions of our corpus, marginal quality gains from strict filtering do not justify large data loss. In our case, the best compromise is Demucs with the most permissive threshold (lowest filter cutoff), which yields balanced improvements across metrics. We discard the no-filtering option, since even permissive thresholds reduce metric variance by removing extreme-condition cases.

\section{Limitations and Future Work}
\label{sec:lim}
Although our work provides a testbed for evaluating preprocessing pipelines without training TTS systems, it remains essential to quantify how objective improvements in dataset quality relate to downstream synthesis performance. We plan to measure the correlation between the composite score and TTS outcomes by training representative TTS models on metric-selected subsets.


From an operational standpoint, the current low-cost, CPU-friendly pipeline is constrained by the STT stage: obtaining accurate, CPU-efficient transcriptions remains a bottleneck. Future work will explore alternative STT models to improve the accuracy–latency trade-off and will implement a lightweight speaker diarization model. After integrating these components, the \emph{in-the-wild} Argentine Spanish dataset will be prepared for public release.

\section{Conclusions}
\label{sec:conclusions}


We introduced a reproducible, metric-driven methodology to evaluate preprocessing pipelines for \emph{in-the-wild} TTS corpora. Experiments applying a low-cost, CPU-friendly processing chain to the first \emph{in-the-wild} Argentine Spanish dataset enabled systematic comparison of 24 pipeline configurations and exposed clear trade-offs between dataset size, signal quality, acoustic conditions and speech-preservation. 

Empirically, Demucs-based denoising with permissive filtering provided the best overall compromise for our testbed, although optimal settings depend on the target weighting of the evaluation criteria. The proposed methodology allows selecting and optimizing pipeline configurations without training a TTS model for each candidate subset, thereby accelerating development and enabling faster, comparable and more cost-effective preprocessing for low-resource settings.


\begin{table}[t]
\centering
\caption{Configuration scores for each metric category.}
\vspace{-6pt}
\label{tab:config-scores}
\resizebox{\columnwidth}{!}{%
\begin{tabular}{@{} l c c c c c @{}}
\toprule
\textbf{Config} &
\textbf{DR}  &
\textbf{SQ}  &
\textbf{AP}  &
\textbf{SD}  &
\textbf{Total} \\
\midrule
Demucs + DNSMOS: 2.7 & \textbf{0.02} & 2.30 & 1.29 & 0.48 & \textbf{4.08} \\
\addlinespace[1pt]
DFN + NISQA: 3 & 0.15 & 2.67 & 1.37 & 0.63 & 4.83 \\
\addlinespace[1pt]
No-den + DNSMOS: 2.7 & 0.19 & 2.85 & 1.82 & \textbf{0.07} & 4.93 \\
\addlinespace[1pt]
Demucs + DNSMOS: 3.4 & 0.8 & \textbf{2.23} & \textbf{1.24} & 0.78 & 5.05 \\
\addlinespace[1pt]
No-den + NISQA: 4.2  & 0.89 & 2.53 & 1.65 & 0.46 & 5.53 \\
\bottomrule
\end{tabular}%
}
\end{table}

\vfill\pagebreak

\bibliographystyle{IEEEbib}
\bibliography{refs}


\end{document}